\documentclass[aps,prd,reprint,amsmath,amssymb,superscriptaddress,floatfix,twocolumn]{revtex4}
\usepackage{graphicx}
\usepackage{amsfonts}
\usepackage{verbatim}
\usepackage{amsmath}
\usepackage{subfigure}
\usepackage{tikz,mathpazo}
\usepackage{amssymb}
\usepackage{rotating}
\usepackage{booktabs}
\usepackage{xcolor}
\usepackage{soul}
\usepackage{color}
\usepackage{slashed}
\usepackage{multirow}
\usepackage{makecell}
\usepackage{epsf}
\usepackage{ulem}
\usepackage{cancel}
\usepackage{color,bm}
\usepackage{bm}
\usepackage{mathtools}

\usepackage[colorinlistoftodos]{todonotes}

\usepackage[colorlinks=true,citecolor=cyan,urlcolor=blue,bookmarks=true,bookmarks=true,bookmarksopen=true,bookmarksnumbered=true,bookmarksopenlevel=3]{hyperref}

\usepackage[compat=1.1.0]{tikz-feynman}

\definecolor{airforceblue}{rgb}{0.36, 0.54, 0.66}
\definecolor{steelblue}{rgb}{0.27, 0.51, 0.71}
\definecolor{amber}{rgb}{1.0, 0.49, 0.0}

\allowdisplaybreaks[4]

\begin{document}

\title{Triangle mechanism in the decay process $J/\psi \to K^- K^+ a_1(1260)$}
\author{Xuan Luo}
\author{Dazhuang He}
\author{Yiling Xie}
\author{{Hao Sun}\footnote{Corresponding author: haosun@mail.ustc.edu.cn \hspace{0.2cm} haosun@dlut.edu.cn}}
\affiliation{ Institute of Theoretical Physics, School of Physics, Dalian University of Technology, \\ No.2 Linggong Road, Dalian, Liaoning, 116024, P.R.China }
\date{\today}

\begin{abstract}  
The role of triangle mechanism in the decay process $J/\psi \to K^- K^+ a_1(1260)$ is probed. In this mechanism, a close-up resonance with mass $1823$ MeV and width $122$ MeV decays into $K^* \phi, K^* \to K \pi$ and then $K^* \bar{K}$ fuses into the $a_1(1260)$ resonance. We find that this mechanism leads to a triangle singularity around $M_{\rm inv}(K^- a_1(1260))\approx 1920$ MeV, where the axial-vector meson $a_1(1260)$ is considered as a dynamically generated resonance. With the help of the triangle mechanism we find sizable branching ratios $\text{Br}(J/\psi \to K^- K^+ a_1(1260),a_1 \to \pi \rho)=1.210 \times 10^{-5}$ and $\text{Br}(J/\psi \to K^- K^+ a_1(1260))=3.501 \times 10^{-5}$. Such a effect from triangle mechanism of the decay process could be investigated by such as BESIII, LHCb and Belle-II experiments. This potential investigation can help us obtain the information of the axial-vector meson $a_1(1260)$. 

\vspace{0.5cm}
\end{abstract}
\maketitle
\setcounter{footnote}{0}

\section{Introduction}
\label{I}

Researchers have shown an increased interest in triangle singularities which were first coped with by Landau \cite{Karplus:1958zz,Landau:1959fi} in 1960s.
A considerable amount of literature \cite{Peierls:1961zz,Aitchison:1964zz,Bronzan:1964zz,Coleman:1965xm,Schmid:1967ojm} has been published on triangle singularities which are essentially brought about triangle loop Feynman diagrams where an external particle 1 decays into A and B particles, internal particle B decays into particle
C and an external particle 2, and then particles A and C fuse into an external particle 3. To produce triangle singularities,  according to Coleman-Norton Theorem \cite{Coleman:1965xm}, the process can occur classically and then all three intermediate particles must be put on shell and be collinear simultaneously. If there are zero width for all the internal particles, the loop integral turns out to be infinite (see e.g. \cite{Guo:2019twa}). Nevertheless, particle B has a finite width since it can decay to particle C and 2, which leads to a finite peak in the invariant mass distributions. This peak can be accessed in experiments.
Instead of evaluating the whole amplitude of a Feynman diagram including triangle loop, there is a more simple and practical way addressed in Ref.~\cite{Bayar:2016ftu} to find the position of a triangle singularity. The condition for producing a triangle singularity is just $q_{on}=q_{a_-}$, where $q_{on}$ is the on shell momentum of particle A or B in the particle 1 rest frame and $q_{a_-}$ defines one of the two solutions for the momenta of particle B when B, C are on shell to produce particle 3. Previous research \cite{Bayar:2016ftu} has established a convenient way to handle triangle mechanism method. A considerable amount of literature has been published on triangle mechanism. Searching a reaction showing a peak due to the triangle singularity fails at the beginning. In 2015, the COMPASS Collaboration reported a peak in 1420 MeV for invariant mass of final state $\pi f_0(980)$ \cite{Adolph:2015pws}. Soon the peak was explained as a triangle singularity corresponding to $\pi f_0(980)$ decay mode of $a_1(1260)$ resonance \cite{Liu:2015taa,Ketzer:2015tqa,Aceti:2016yeb}. Another consideration of triangle mechanism lies to the abnormally enhanced of isospin violating process $\eta(1405) \to \pi f_0(980)$ compared to the process $\eta(1405) \to \pi a_0(980)$  \cite{BESIII:2012aa}. This abnormally enhancement then be suggested due to the triangle singularities  \cite{Wu:2011yx,Aceti:2012dj,Wu:2012pg,Achasov:2015uua,Achasov:2018swa}. Also, to clarify an enhancement in the $K \Lambda(1405)$ invariment mass distribution of the $\gamma p \to K \Lambda(1405)$ at about $\sqrt{s}=2110$ MeV \cite{Moriya:2013hwg}, the authors of Ref.~\cite{Wang:2016dtb} tied the peak to a triangle singularity coming from a resonance $N^*(2030)$ dynamically generated from the vector-baryon interaction. In addition, there are a lot of research examples where the triangle mechanism plays an important role \cite{Xie:2016lvs,Roca:2017bvy,Debastiani:2017dlz,Samart:2017scf,Sakai:2017hpg,Pavao:2017kcr,Sakai:2017iqs,Xie:2017mbe,Bayar:2017svj,Dai:2018hqb,Xie:2018gbi,Liang:2019jtr,Liu:2019dqc,Jing:2019cbw,Nakamura:2019nwd,Sakai:2020fjh,Molina:2020kyu,Sakai:2020ucu,Debastiani:2018xoi,Oset:2018zgc,Dai:2018rra,Dai:2018zki}.

In Ref.~\cite{Debastiani:2016xgg}, the $K^* \bar{K}$ peak related to the $a_1(1420)$ demonstrates that $f_1(1285)$ can decay into the $K^*\bar{K}+c.c.$, and there is a triangle singularity enhanced decay mode $\pi a_0(980)$ for the $f_1(1285)$ where $f_1(1285) \to K^* \bar{K}$, $K^*\to \pi K$ and then $K\bar{K} \to a_0(980)$.
In this paper, we focus on the reaction process $J/\psi \to K \bar{K} a_1(1260), a_1 \to \pi^+ \rho^-$, where $a_1(1260)$ is viewed as a dynamically generated resonance through using the chiral unitary approach \cite{Roca:2005nm,Lutz:2003fm}, i.e. it can be described as a quasi bound state of dihadron in coupled channels.
The $a_1$(1260) has been probed in the radiative decay process which is viewed as dynamically generated hadron state \cite{GomezDumm:2003ku,Wagner:2008gz,Dumm:2009va}. The $a_1(1260)$ resonance is investigated in a three-body $\tau$ lepton decay process where the triangle mechanism play a important role. Recently, the authors in Ref.~\cite{Zhang:2018kdz} probe the strengths of the $a_1(1260)$ photoproduction in the $\gamma p \to a_1(1260)^+ n$ and $\gamma p \to \pi^+ \pi^+ \pi^- n$ reactions via the $\pi$-exchange mechanism.
We reach a peak of the invariant mass $M_{\rm inv}(K^+ a_1)$ at around 1920 MeV by applying triangle mechanism, where a close-up dynamically generated resonance decays into $K^* \phi$, $K^* \to K \pi$ and then $K^* \bar{K}$ fuses into the $a_1(1260)$ resonance. 
We apply the experimental data of the branching ratio of the decay $J/\psi \to \bar{K} K^{*} \phi$ to determine the coupling strength of the $J/\psi \bar{K} K^* \phi$ vertex.
For the $a_1 K^- K^{*+}$ vertex inside the triangle loop of the decay process, we apply the chiral unitary approach by viewing the $a_1(1260)$ as dynamically generated hadron state. The branching ratio of the underlying decay process is obtained.
Similarly, a research \cite{Xie:2016lvs} predicted a $f_2(1810)$ triangle singularity, coming from a nearby $f_2(1640)$ going into $K^* \bar{K}^*$, $K^* \to K \pi$, followed by $K^* \bar{K}$ fusing into $a_1(1260)$.
The $J/\psi \to K \bar{K} a_1(1260)$, $a_1 \to \pi^+ \rho^-$ process we suppose is a practical example of a physical process where triangle mechanism can work. We also perform a quantitative calculation of the triangle loop amplitude $t_T$.

This paper has been divided into four sections. The Section \ref{II} deals with the calculation framework and formalism for working out the decay amplitude of $J/\psi \to K \bar{K} a_1(1260)$, $a_1 \to \pi^+ \rho^-$ including triangle mechanism. Among this section, the vertex coupling involved in the tree level process $J/\psi \to K^* \bar{K} \phi$ has been calculated where we have introduced a dynamically generated resonance propagator. In Section \ref{III}, we give out the numerical results related to the triangle singularity around 1920 MeV and then make a discussion about them. Also, the corresponding decay branching ratios have been obtained. We reach our conclusion in Section \ref{IV}.

\section{FORMALISM}
\label{II}

\begin{figure}[h]
\begin{center} 
\subfigure[]{
\begin{tikzpicture}
	\begin{feynman}
	\vertex(a1){\( J/\psi \)};
	\vertex[right=1.6cm of a1] (a2);
	\vertex[right=1cm of a2] (a3);
	\vertex[right=0.6cm of a3] (a4);
	\vertex[right=0.8cm of a4] (a5);
	\vertex[right=1.5cm of a5] (a6);
	
	\vertex[above=1.2cm of a3](u1);
	
	\vertex[below=1.5cm of a4](d1);
	\vertex[right=1.2cm of d1](d0);
	\vertex[below=0.5cm of d0](d2);
	\vertex[right=1.2cm of d2](d3);	
	
	\vertex[above=0.8cm of d3](d31);
	\vertex[below=0.5 of d3](d32);

	\diagram*[baseline=(a3.base)] {
	{
	 (a1) --[fermion] (a2)--[fermion, edge label=\( \phi(P-q) \)](a5)--[fermion,edge label=\( K^- \)](a6),
	},
    (a2) --[fermion,edge label=\( K^+ \)] (u1),
    (a5) --[fermion,edge label=\( K^+(P-q-k) \)] (d1),
	(a2)--[fermion,edge label'=\( K^{*-}(q) \)](d1)--[very thick,edge label'=\( a_1(1260) \)](d2),
	(d2) --[fermion,edge label=\( \rho^- \)] (d31),
	(d2) --[fermion,edge label=\( \pi^+ \)] (d32),
	};
	\end{feynman}
\end{tikzpicture}
}
\subfigure[]{
\begin{tikzpicture}
	\begin{feynman}
	\vertex(a1){\( J/\psi \)};
	\vertex[right=1.6cm of a1] (a2);
	\vertex[right=1cm of a2] (a3);
	\vertex[right=0.6cm of a3] (a4);
	\vertex[right=0.8cm of a4] (a5);
	\vertex[right=1.5cm of a5] (a6);
	
	\vertex[above=1.2cm of a3](u1);
	
	\vertex[below=1.5cm of a4](d1);
	\vertex[right=1.2cm of d1](d0);
	\vertex[below=0.5cm of d0](d2);
	\vertex[right=1.2cm of d2](d3);	
	
	\vertex[above=0.8cm of d3](d31);
	\vertex[below=0.5 of d3](d32);

	\diagram*[baseline=(a3.base)] {
	{
	 (a1) --[fermion] (a2)--[fermion, edge label=\( \phi(P-q) \)](a5)--[fermion,edge label=\( K^+ \)](a6),
	},
    (a2) --[fermion,edge label=\( K^- \)] (u1),
    (a5) --[fermion,edge label=\( K^-(P-q-k) \)] (d1),
	(a2)--[fermion,edge label'=\( K^{*+}(q) \)](d1)--[very thick,edge label'=\( a_1(1260) \)](d2),
	(d2) --[fermion,edge label=\( \rho^- \)] (d31),
	(d2) --[fermion,edge label=\( \pi^+ \)] (d32),
	};
	\end{feynman}
\end{tikzpicture}
}
\caption{The Feynman diagrams of the decay process $J/\psi \to K^- K^+ a_1(1260), a_1 \to \pi^+ \rho^-$ involving a triangle loop.}
\label{fig1}
\end{center}
\end{figure}
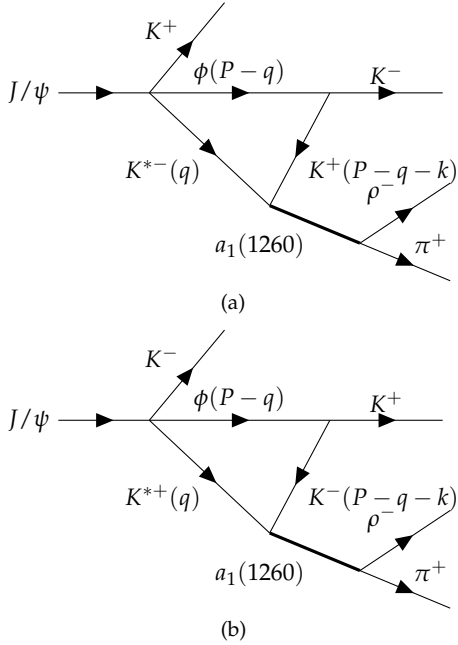
We plot the Feynman diagrams of the decay process $J/\psi \to K^- K^+ a_1(1260)$ involving a triangle loop in Fig.\ref{fig1}, where the meson $J/\psi$ first decays into two vector mesons $\phi$, $K^{*}$ and a pesudoscalar meson $\bar{K}$, and then the meson $\phi$ is converted into $K$ and $\bar{K}$. The $\bar{K}$ can move faster than $ K^{*}$, so they can combine to generate the $a_1(1260)$.  Finally, we consider that the $a_1(1260)$ continues to decay into $\pi^+ \rho^-$ final states.  

We take Fig.\ref{fig1}(b) for example to perform the following discussion since Fig.\ref{fig1}(a) and (b) have nearly the same amplitude.
In order to find the position of triangle singularity in complex-$q$ plane, analogously to Ref.~\cite{Bayar:2016ftu, Wang:2016dtb, Huang:2020kxf}, we use
\begin{equation}\label{eq1}
q_{+}^{\text{on}}=q_-^{\text{a}},\quad \text{and} \quad q_{+}^{\text{on}}=\frac{\lambda^{\frac{1}{2}}(s,M_\phi^2,M_{K^{*+}}^2)}{2\sqrt{s}},
\end{equation}
where the $q_{+}^{\text{on}}$ is the on shell three momentum of the $K^{*+}$ in the center of mass frame of $\phi K^{*+}$, $s$ denotes the squared invariant mass of $\phi$ and $K^{*+}$, and $\lambda(x, y, z)=x^2 +y^2 +z^2 -2xy-2yz -2xz$ is the K\"ahlen function.

Meanwhile, $q_-^{\text{a}}$ can be obtained by analyzing the singularity structure of the triangle loop, which is given by
\begin{equation}\label{eq2}
\begin{aligned}
q_-^{\text{a}} =\gamma(\nu E_{K^{*+}}-p_{K^{*+}}^*)-i \epsilon
\end{aligned}	
\end{equation}
with definition
\begin{equation}\label{eq3}
\begin{aligned}
\nu =& \frac{k}{E_{a_1}}, \qquad \qquad \qquad \gamma=\frac{1}{\sqrt{1-\nu^2}}=\frac{E_{a_1}}{m_{a_1}},\\
E_{K^{*+}}=&\frac{m_{a_1}^2+m_{K^{*+}}^2-m_{k^-}^2}{2m_{a_1}},\ p_{K^{*+}}^*=\frac{\lambda^{\frac{1}{2}}(m_{a_1}^2,M_{K^-}^2,M_{K^{*+}}^2)}{2m_{a_1}}.
\end{aligned}
\end{equation}
where $E_{K^{*+}}$  and $p_{K^{*+}}^*$ are the energy and momentum of the $K^{*+}$ meson in the center of mass frame of the $K^{*+}\phi$ system, $\nu$ and $\gamma$ are the velocity of the $a_1$ and Lorentz boost factor, respectively. In addition,
\begin{equation}\label{eq4}
\begin{aligned}
E_{a_1}=\frac{s+m_{a_1}^2-m_{K^-}^2}{2\sqrt{s}}, \qquad k=\frac{\lambda^{\frac{1}{2}}(s,m_{a_1}^2,m_{K^-}^2)}{2\sqrt{s}}. 
\end{aligned}
\end{equation}
When Eq.(\ref{eq1}) is established, we need to consider the case that all three intermediate particles in the triangle loop are on shell and the angular $z$ between momentum $q$ and $k$ is taken as $z=-1$, i.e. the momentum of particle $K^{*+}$ is anti-parallel to the momentum of the $a_1(1260)$ in the $K^{*+}K^-$ center of mass frame. Now, by letting the mass of the $a_1(1260)$ slightly larger than the mass sum of $K^*$ and $K$ mesons and applying Eq.(\ref{eq1}), one can find a triangle singularity at around $\sqrt{s} = 1920$ MeV keeping $K^{*+}=891.66$ MeV and $K^-=493.68$ MeV~\cite{Tanabashi:2018oca} in mind. If we use in Eq.\eqref{eq1} complex masses $(M-i\Gamma/2)$ of vector mesons which include widths of $K^*$ and $\phi$ mesons, the solution of Eq.(\ref{eq1}) is then $(1920-48i)$ MeV. This solution implies that the triangle singularity has a "width" 96 MeV.

\subsection{The decay process $J/\psi \to K^- \phi K^{*+}$}
Before writing the whole amplitude of the Feynman diagram in FIG.\ref{fig1}(b), the generalized vertex $V_{J/\psi, K^- \phi K^{*+}}$ needs to be calculated firstly. 
Experimentally we have the $K^* \phi$ invariant mass distribution in Fig.10(a) of Ref.\cite{Ablikim:2007ev}. One can find a broad peak around 1800 MeV in $K^* \phi$ invariant mass distribution. This structure indicates that there should better be a form factor in the one to three amplitude, which comes from the interaction of $K^*\phi$ to give a resonance around 1800 MeV. This structure is somewhat important since if we do not add this structure in the one to three vertex, after carefully calculation one can not find a clear singularity in $K^+ a_1(1260)$ invariant mass distribution of $J/\psi \to K^- K^+ a_1(1260)$ decay process. In that case, the clear peak of $K^+ a_1(1260)$ in $|t_T|^2$ will be expunged by the phase space and kinematic factors in the total $J/\psi \to K^- K^+ a_1(1260)$ decay process.

Now we introduce some kind of propagator $X$ that can decay into $K^* \phi$. The process then becomes $J/\psi \to K^- X \to K^- \phi K^{*+}$. Under conservation of Strangness, Isospin and Spin for $X \to K^* \phi$, the low lying vector meson $X$ should satisfy Strangness=0, Isospin=1/2 and Spin=1. In Refs. \cite{Geng:2008gx,GarciaRecio:2010ki}, the vector meson $K_1(1650)$ is regarded well as dynamically generated state from the vector vector interaction which corresponds to the pole position (1665,-95). It is supposed to couple with two vector mesons such as $K^{*} \rho, K^{*} \omega, K^{*} \phi$. However, the width of $K_1(1650)$ is not very large and the mass of $K_1(1650)$ is somewhat far away from the $K^* \phi$ threshold which results in a small possibility of decay from $K_1(1650)$ to $K^{*} \phi$ final states. Alternatively, a pole position (1823,-61) was reported by applying a different subtraction constants $a=-3.1$ \cite{GarciaRecio:2010ki}. This potential "$K_1$" state has a suitable width and its mass 1823 MeV is close to the $K^* \phi$ threshold. Also, its coupling from $K^* \phi$ shown in Ref.~\cite{GarciaRecio:2010ki} is relatively large. On basis of the above considerations, we choose this reported pole of $K_1$ type as the propagator X, which has a resonant shape
\begin{equation}
\begin{aligned}
F(M_{K^* \phi})=\frac{M_X \Gamma_{M_X}}{M_{K^* \phi}^2-M_{X}^2+i M_{X}\Gamma_{M_X}},
\end{aligned}
\end{equation}
where $M_X,\Gamma_{M_X}$ are taken by 1823 MeV and 122 MeV, respectively.
The $J/\psi \to K^- \phi K^{*+}$ decay amplitude can then be written as
\begin{equation}\label{amp1}
\begin{aligned}
-it_{J/\psi, K^- \phi K^{*+}} &=-i\mathcal{C} \varepsilon_{ijk} \epsilon_{i}(J/\psi)\epsilon_{j}(\phi)\epsilon_{k}(K^*) F(M_{K^{*+} \phi}).\\
\end{aligned}
\end{equation}
The coefficient $\mathcal{C}$ is obtained by comparing the calculated $J/\psi \to\phi K^- K^{*+}$ decay branching ratio with those from experiment. Note that the amplitude in Eq.\eqref{amp1} is only a bit rough, since we have neglected all the other contributions to the process, such as resonances that couple to each pair of the three final state mesons. In Refs.\cite{Ablikim:2007ev,Zyla:2020zbs}, the decay branching ratio for $J/\psi \to\phi K^* \bar{K} +c.c.$ is $(2.18 \pm 0.23)\times 10^{-3}$. Here, we only focus on the 
$J/\psi \to\phi K^- K^{*+}$ process and the relation between above two decay branching ratio is
\begin{equation}
\begin{aligned}
&Br(J/\psi \to\phi K^- K^{*+}) =\frac{1}{4}Br(J/\psi \to\phi K^* \bar{K} +c.c.).
\end{aligned}
\end{equation}

The differential decay width over the invariant mass distribution $K^{*+}\phi$ can be written as
\begin{equation}\label{1to3}
\begin{aligned}
&\frac{d\Gamma_{J/\psi \to\phi K^- K^{*+}}}{dM_{inv}(K^{*+} \phi)}=\frac{1}{(2\pi)^5}\frac{|\vec k_{K^{*+}}^* | |\vec k_{K^-}|}{16m_J^2}
\\
&\cdot \overline{\sum} 
|t_{J/\psi \to\phi K^- K^{*+}}|^2d\Omega_{K^{*+}}^*d\Omega_{K^-},
\end{aligned}
\end{equation}
where $m_J$ is the mass of $J/\psi$, $|\vec k_{K^{*+}}^* |$ and $\Omega_{K^{*+}}^*$ are the absolute value of the $K^{*+}$ three momentum and the $K^{*+}$ solid angle in the center of mass frame of the final $K^{*+}\phi$ system, respectively. Whereas, $|\vec k_{K^-} |$ and $\Omega_{k_{K^-}}$ are the absolute value of the $K^-$ three momentum and the $K^-$ solid angle in the rest frame of the initial $J/\psi$ meson, respectively.
To perform the calculation in Eq.\eqref{1to3}, we use the polarization summation formula:
\begin{equation}\label{pol}
\begin{aligned}
\sum_{\rm pol} \varepsilon_\mu(p) \varepsilon_\nu(p)=-g_{\mu\nu}+\frac{p_\mu p_\nu}{m^2},
\end{aligned}
\end{equation}
Then one can obtain
\begin{equation}
\begin{aligned}
&\frac{\mathcal{C}^2}{\Gamma_{J/\psi}}=\frac{Br(J/\psi \to\phi K^- K^{*+}) }{\int dM_{inv}(K^{*+}\phi)\frac{d\Gamma_{J/\psi \to\phi K^- K^{*+}}}{dM_{inv}(K^{*+}\phi)}}.
\end{aligned}
\end{equation}

\subsection{The role of triangle mechanism in the decay $J/\psi \to K^- K^+ a_1(1260),a_1 \to \pi^+ \rho^-$}

In the previous subsection we have calculated the transition strength of the decay process $J/\psi \to K^- \phi K^{*+}$. 
Now we focus on the triangle diagram amplitudes required by $J/\psi \to K^- K^+ a_1(1260)$, $a_1 \to \pi^+ \rho^-$ process. The feynman diagrams are shown in Fig.\ref{fig1} where $J/\psi$ decays into $\phi K^* \bar{K}$, the $\phi$ decays into $\bar{K} K$, and then the $K^*$ and $\bar{K}$ fuse into the $a_1(1260)$. The particles' ID and momentum informations are labelled in the diagrams. 
This triangle mechanism can take place as long as the $a_1(1260)$ couples to the $K^*\bar{K}$ pair, thus the triangle singularity induced decay process can be used to gain valuable information for the $a_1(1260)$. 
Finally, the $a_1(1260)$ decays into $\pi^+ \rho^-$. At the beginning, we need to evaluate the $\phi K^+ K^-$ vertex for Fig.\ref{fig1}(b) which can be obtained from the vector-pseudoscalar-pseudoscalar Lagrangian
\begin{equation}\label{eq5}
\begin{aligned}
\mathcal{L}_{VPP}=&-ig \langle V^\mu [P,\partial_{\mu} P] \rangle,
\end{aligned}
\end{equation}
where the $\langle \  \rangle$ represents the SU(3) trace. The coupling constant $g$, vector meson mass, and the decay constant of pion are taken as \cite{Pich:1995bw}
\begin{equation}\label{eq6}
\begin{aligned}
g=\frac{M_V}{2f_\pi}, \qquad M_V=800 \ \text{MeV}, \qquad f_{\pi} =93 \ \text{MeV}.
\end{aligned}
\end{equation} 
The $V$ and $P$ in Eq.(\ref{eq5}) are the  vector meson matrix and pseudoscalar meson matrix in the SU(3) group, respectively \cite{Liang:2017ijf}
\begin{equation}\label{matrix}
\begin{aligned}
&P=\begin{pmatrix}
\frac{\pi^0}{\sqrt{2}}+\frac{\eta}{\sqrt{3}}+\frac{\eta\prime}{\sqrt{6}} &\pi^+ &K^+\\
 \pi^-&-\frac{\pi^0}{\sqrt{2}}+\frac{\eta}{\sqrt{3}}+\frac{\eta\prime}{\sqrt{6}} &K^0\\
 K^-& \bar{K}^0& -\frac{\eta}{\sqrt{3}}+\sqrt{\frac{2}{3}}\eta\prime
\end{pmatrix}_, \qquad\\
&V=\begin{pmatrix}
	\frac{\rho^0}{\sqrt{2}}+\frac{\omega}{\sqrt{2}}& \rho^+& K^{*+}\\
	\rho^-&-\frac{\rho^0}{\sqrt{2}}+\frac{\omega}{\sqrt{2}} & K^{*0}\\
	K^{*-}& \bar{K}^{*0}& \phi 
	\end{pmatrix}_.
\end{aligned}
\end{equation}

The total amplitude of the $J/\psi \to K^- K^+ a_1$ decay process as shown in Fig.\ref{fig1}(b) can be written down straightly:
\begin{equation}\label{t1}
\begin{aligned}
-i t =&-i  \mathcal{C} \ F(K^- a_1)  \varepsilon_{ijk} \varepsilon_i(J/\psi)\varepsilon_j(\phi)\varepsilon_k(K^*) 
\\
&\cdot \int \frac{d^4 q}{(2\pi)^4} \frac{i}{q^2-m_{K^{*+}}^2+im_{K^*}\Gamma_{K^*}}
\\
&\cdot \frac{i}{(P-q)^2-m_\phi^2+im_{\phi}\Gamma_{\phi}}\frac{i}{(P-q-k)^2-m_{K^-}^2+i\varepsilon}
\\
&\cdot  (-ig)(p_{K^+}-p_{K^-})_\mu \varepsilon^\mu(\phi) (-ig_{a_1,K^{*+}K^-})\varepsilon(K^*) \cdot \varepsilon(a_1),
\end{aligned}
\end{equation}
where the $g_{a_1,K^{*+}K^-}$ is the coupling of the $a_1(1260)$ to $K^{*+}K^-$, and $P^0=M_{\rm inv}(K^+ a_1)$ in the $K^+ a_1$ rest frame. 
We have assumed that only the spatial components of the polarization vector of vector mesons are nonvanishing, which leads to the vanishing zero component of the polarization vector and the completeness relation for the polarization vectors written as
\begin{align}\label{pol1}
\sum_{\rm pol} \varepsilon_\mu(p) \varepsilon_\nu(p)=\delta_{\mu\nu}+\frac{p_\mu p_\nu}{m^2},
\end{align}
where $i,j$ is lorentz indices from 1 to 3.
In Eq.~\eqref{t1}, after integrating over $\vec{q}$ only the vector $\vec{k}$ remains, therefore for a function $f(\vec{q}, \vec{k})$ we have
\begin{equation}\label{inte}
\begin{aligned}
\int d^3 \vec{q} \; q_i \; f(\vec{q}, \vec{k}) =A\;  k_i,\\
A=\int d^3 \vec{q} \frac{\vec{q}\cdot \vec{k}}{|\vec{k}|^2}\; \;f(\vec{q}, \vec{k}).
\end{aligned}
\end{equation}
Considering Eqs.~\eqref{pol1} and \eqref{inte}, Eq.~\eqref{t1} can be simplified as
\begin{equation}\label{t2}
\begin{aligned}
t &=\mathcal{C} \ F \ \varepsilon_{ijk} \varepsilon_i(J/\psi)\varepsilon_k(a_1) i \int \frac{d^4 q}{(2\pi)^4} \frac{1}{q^2-m_{K^{*+}}^2+im_{K^*}\Gamma_{K^*}}\\
&\frac{(2k+q)_j \ g \ g_{a_1,K^{*+}K^-}}{(P-q)^2-m_\phi^2+im_{\phi}\Gamma_{\phi}}\frac{1}{(P-q-k)^2-m_{K^-}^2+i\varepsilon}\\
&= i\mathcal{C}  F(M_{\rm inv}(K^+ a_1))  g \ g_{a_1,K^{*+}K^-}  \varepsilon_{ijk} \varepsilon_i(J/\psi)\varepsilon_k(a_1) k_j t_T.
\end{aligned}
\end{equation} 
where the zero component of $q$ integration in Eq.~\eqref{t2} has been performed analytically by residue theory, leads to a three-dimensional loop intergral $t_T$ which can be integrated numerically
\begin{equation}\label{tT}
\begin{aligned}
 t_T &= \int \frac{{\rm d}^3 q}{(2\pi)^3} \; \frac{1}{8\, \omega_{K^-}\, \omega_{\phi} \, \omega_{K^{*+}}} \; \frac{1}{k^0-\omega_{K^-}- \omega_{\phi} } \; \\
 &\times \frac{1}{M_{\rm inv}(K^+ a_1) + \omega_{K^{*+}} +\omega_{K^-} -k^0}\;  \\
   & \times \, \frac{1}{M_{\rm inv}(K^+ a_1) - \omega_{K^{*+}}- \omega_{K^-} -k^0 +i\frac{\Gamma_{K^{*+}}}{2}}  \\
  &\times  \bigg[ \frac{2 M_{\rm inv}(K^+ a_1)  \omega_{K^{*+}}+2k^0 \omega_{K^-} }{M_{\rm inv}(K^+ a_1)-\omega_{\phi}-\omega_{K^{*+}} +i\frac{\Gamma_{\phi}}{2}+i\frac{\Gamma_{K^{*+}}}{2}} \\
  &-\frac{2(\omega_{K^{*+}}+\omega_{K^-})(\omega_{K^{*+}}+\omega_{\phi}+\omega_{K^-})}{M_{\rm inv}(K^+ a_1)-\omega_{\phi}-\omega_{K^{*+}} +i\frac{\Gamma_{\phi}}{2}+i\frac{\Gamma_{K^{*+}}}{2}} \bigg] \left(  2+ \frac{\vec q \cdot \vec k}{\left. {\vec k} \right. ^2} \right),
\end{aligned}
\end{equation} 
with 
\begin{equation}
\begin{aligned}
 &\omega_{\phi}=\sqrt{\vec q\,^2 + m^2_{\phi}},\quad \omega_{K^-}=\sqrt{(\vec q + \vec k)^2 + m^2_{K^-}},\\
 &\omega_{K^{*+}}=\sqrt{\vec q\,^2 + m^2_{K^{*+}}},\quad k^0=\frac{M^2_{\rm inv}(K^+ a_1)+m^2_{K^+}-m^2_{a_1}}{2M_{\rm inv}(K^+ a_1)},\\
 &| \vec k |=\frac{\lambda^{1/2} \left( M^2_{\rm inv}(K^+ a_1),\, m^2_{K^+}, \,m^2_{ a_1}\right)}{2M_{\rm inv}(K^+ a_1)}.
 \end{aligned}
\end{equation}
The width of vector mesons $K^{*+}$ and $\phi$ are taken as $\Gamma_{K^{*+}}=48$ MeV and $\Gamma_{\phi}=4.25$ MeV, respectively.
The above integral is regularized with a cutoff $q_{max}$ on the loop integral $|\vec q|$. We take $q_{max}=950$ as in~\cite{Molina:2016pbg} to produce $a_1(1260)$ in the chiral unitary approach.
In the $|\vec t|^2$ level, we need to sum over the spin structure of the external vector mesons. Applying Eq.~\eqref{pol}, we have
\begin{equation}
\begin{aligned}
 \overline{ \sum\limits_{\rm pol}} &\varepsilon_{ijk} \varepsilon_i(J/\psi)\varepsilon_k(a_1) k_j \varepsilon_{\mu\nu\rho} \varepsilon_\mu(J/\psi)\varepsilon_\rho(a_1) k_\nu =\\
 & \frac{1}{3}\left(\frac{2\vec p^2 \vec k^2}{m_{a_1}^2}-\frac{2(\vec p\cdot \vec k)^2}{m_{a_1}^2}+6\vec k^2\right) ,
\end{aligned}
\end{equation}
where $\vec{p}=\vec{P}-\vec{k}$ is the three vector of the final $a_1(1260)$ meson, and the mass of the $a_1(1260)$ is taken as 1230 MeV from PDG~\cite{Tanabashi:2018oca}. Then the distribution of invariant mass $M_{\rm inv}(K^+ a_1)$ in the decay $J/\psi \to K^- K^+ a_1(1260)$ can be written as
\begin{equation}
\begin{aligned}
&\frac{1}{\Gamma_{J/\psi}}\frac{d\Gamma_{J/\psi \to K^- K^+ a_1(1260)}}{dM_{\rm inv}(K^+ a_1)}=\frac{1}{(2\pi)^3}\frac{|\vec{\tilde{q}}_{K^{+}}| \ |\vec p_{K^-}| }{4M^2}\overline{\sum} |t|^2,
\end{aligned}
\end{equation}
where $M$ is the mass of the $J/\psi$ meson.
The three momentums $|\vec{\tilde{q}}_{K^{+}}|$ and $|\vec p_{K^-}|$ in Eq.~\eqref{Gamma} are given by 
\begin{equation}
\begin{aligned}
|\vec{\tilde{q}}_{K^{+}}|&=\frac{\lambda^{1/2}(M_{\rm inv}^2(K^+ a_1),m^2_{K^{+}},m^2_{a_1})}{2M_{\rm inv}(K^+ a_1)},\\
|\vec p_{K^-}|&=\frac{\lambda^{1/2}\left( M^2,m_{K^-}^2,M_{\rm inv}^2(K^+ a_1) \right)}{2M}.
\end{aligned}
\end{equation}
Then the differential branching ratio of the decay process $J/\psi \to K^- K^+ a_1(1260)$ can be written as 
\begin{equation}\label{gamma}
\begin{aligned}
&\frac{1}{\Gamma_{J/\psi}}\frac{d\Gamma_{J/\psi \to K^- K^+ a_1(1260)}}{dM_{\rm inv}(K^+ a_1)}=
\\
&\frac{1}{(2\pi)^3}\frac{|\vec{\tilde{q}}_{K^{+}}| \ |\vec p_{K^-}| }{4M^2}
 \frac{\mathcal{C}^2}{\Gamma_{J/\psi}} \ F^2(M_{\rm inv}(K^+ a_1)) \ g^2 \ g^2_{a_1,K^{*+}K^-}\\
& \times \frac{1}{3}\left(\frac{2\vec p^2 \vec k^2}{m_{a_1}^2}-\frac{2(\vec p\cdot \vec k)^2}{m_{a_1}^2}+6\vec k^2\right) |t_T|^2,
\end{aligned}
\end{equation}
where the $a_1(1260) \to K^* \bar{K}$ vertex is obtained from the chiral unitary approach of Ref.~\cite{Roca:2005nm} with $g_{a_1,K^{*} K}=2390$ MeV. Furthormore, to perform the numerical calculations, we choose the $z$ axis along the direction of the vector $k$ without loss of generality.

Now we add the final $a_1(1260) \to \pi^+ \rho^-$ decay process to our total amplitude. 
To calculate the $|t|^2$, we need to sum over the spin structure of the external vector mesons. Applying Eq.~\eqref{pol}, we have
\begin{equation}
\begin{aligned}
 \overline{ \sum\limits_{\rm pol}} &\varepsilon_{ijk} \varepsilon_i(J/\psi)\varepsilon_k(\rho) k_j \varepsilon_{\mu\nu\rho} \varepsilon_\mu(J/\psi)\varepsilon_\rho(\rho) k_\nu =\\
 & \frac{1}{3}\left(\frac{2\vec p^{\prime 2} \vec k^2}{m_{\rho}^2}-\frac{2(\vec p^\prime\cdot \vec k)^2}{m_{\rho}^2}+6\vec k^2\right) ,
\end{aligned}
\end{equation}
where $\vec p^\prime$ denotes the three momentum of the final $\rho$ meson, and $m_\rho=782$ MeV denotes the mass of the $\rho$ meson. Then we have
\begin{equation}
\begin{aligned}
\displaystyle \overline{ \sum\limits_{\rm pol}} |t|^2 =& \mathcal{C}^2 \ F^2(M_{\rm inv}(K^+ a_1)) \ g^2 \ g^2_{a_1,\pi^+\rho^-}\\
& \times \frac{1}{3}\left(\frac{2\vec p^{\prime 2} \vec k^2}{m_{\rho}^2}-\frac{2(\vec p^\prime\cdot \vec k)^2}{m_{\rho}^2}+6\vec k^2\right) |t_T|^2.
\end{aligned}
\end{equation}

After applying the calculation details in \cite{Pavao:2017kcr}, one can reach the double differential mass distribution in $M_{\rm inv}(K^+ a_1)$ and $M_{\rm inv}(\pi^+ \rho^-)$
\begin{equation}\label{Gamma}
\begin{aligned}
&\frac{1}{\Gamma_{J/\psi}}\frac{d^2 \Gamma_{J/\psi \to K^- K^+ a_1(1260),a_1\to \pi^+ \rho^-}}{dM_{\rm inv}(K^+ a_1)d M_{\rm inv}(\pi^+\rho^-)}=\\
&\frac{1}{4\pi}\frac{1}{(2\pi)^5}\frac{|\vec p_{K^-}||\vec k| }{4M^2} \frac{\mathcal{C}^2}{\Gamma_{J/\psi}} \ |F(M_{\rm inv}(K^+ a_1))|^2 \ g^2|t_T|^2 |\vec{\tilde{q}}_{\rho}|\\
&\times \frac{1}{3}\left(\frac{2\vec p^{\prime 2} \vec k^2}{m_{\rho}^2}-\frac{2(\vec p^\prime\cdot \vec k)^2}{m_{\rho}^2}+6\vec k^2\right) \cdot |t_{K^{*+} K^- \to \pi^+ \rho^-}|^2 d\Omega_k,
\end{aligned}
\end{equation}
where the coupling $g_{a_1 ,\pi^+ \rho^-}$ has been absorbed into $t_{K^{*+} K^- \to \pi^+ \rho^-}$ denoting the isospin-one amplitude of the final $VP \to VP$ scattering. This amplitude can be calculated by solving the Bethe-Salpeter equation
\begin{equation}
\begin{aligned}
t = \frac{V}{1-V \ G} \ ,
\end{aligned}
\end{equation}
where $G$ is the loop function given in \cite{Roca:2005nm},
and $V$ is a $2\times 2$ matrix of the interation kernel, the two channels are taken by 1 for $K^* K$ and 2 for $\rho\pi$. The corresponding transition potentials come from the Lagrangian involving $VVPP$ coupling under the local hidden gauge approach \cite{Roca:2005nm}
\begin{equation}
\begin{aligned}
V_{ij} =& -\frac{C_{ij}}{8f^2}\bigg[ 3s-(M^2+m^2+M^{\prime 2}+m^{\prime 2})\\
		&-\frac{1}{s}(M^2-m^2)(M^{\prime 2}-m^{\prime 2}) \bigg],
\end{aligned}
\end{equation}
where the $C_{ij}$ are coefficients related to different particles and isospin basis $(S,I)$ \cite{Roca:2005nm}. $M$, $m$ are respectively vector and pseudoscalar mesons in channel $i$, and $M^\prime$, $m^\prime$ are respectively vector and pseudoscalar mesons in channel $j$.
Also, after calculating the corresponding C-G coefficient we have 
\begin{equation}
\begin{aligned}
t_{K^{*+} K^- \to \pi^+ \rho^-}=-\frac{1}{2}t_{K^{*} \bar{K} \to \pi\rho }.
\end{aligned}
\end{equation}
The three momentums $|\vec{\tilde{q}}_{\rho}|$ and $|\vec p_{K^-}|$ in Eq.~\eqref{Gamma} are given by 
\begin{equation}
\begin{aligned}
|\vec{\tilde{q}}_{\rho}|&=\frac{\lambda^{1/2}(M_{\rm inv}^2(\pi^+\rho^-),m^2_{\pi},m^2_{\rho}}{2M_{\rm inv}(\pi^+\rho^-)},\\
|\vec p_{K^-}|&=\frac{\lambda^{1/2}\left( M^2,m_{K^-}^2,M_{\rm inv}^2(K^+ a_1) \right)}{2M}.
\end{aligned}
\end{equation}
In addition, without loss of generality, we choose the four vector $p^\prime$ of the final $\rho^-$ meson as the $z$ axis, then there exists a solid angle integral element $\displaystyle \frac{1}{4\pi}d\Omega_k$ in Eq.~\eqref{Gamma} according to the general three-body phase space integration formalism \cite{Tanabashi:2018oca}. 
Finally, the integration range of $M_{\rm inv}(\pi^+\rho^-)$ is $(m_{\pi^+}+m_{\rho^-}, M_{\rm inv}(K^+ a_1)-m_{K^-})$ as usual and there is also a factor four added in the numerial calculation due to the two Feynman diagram contributions for the underlying process.

\section{Numerical calculation}
\label{III}

\begin{figure*}[htp!]
\begin{center}
\subfigure[]
{
\includegraphics[scale=0.35]{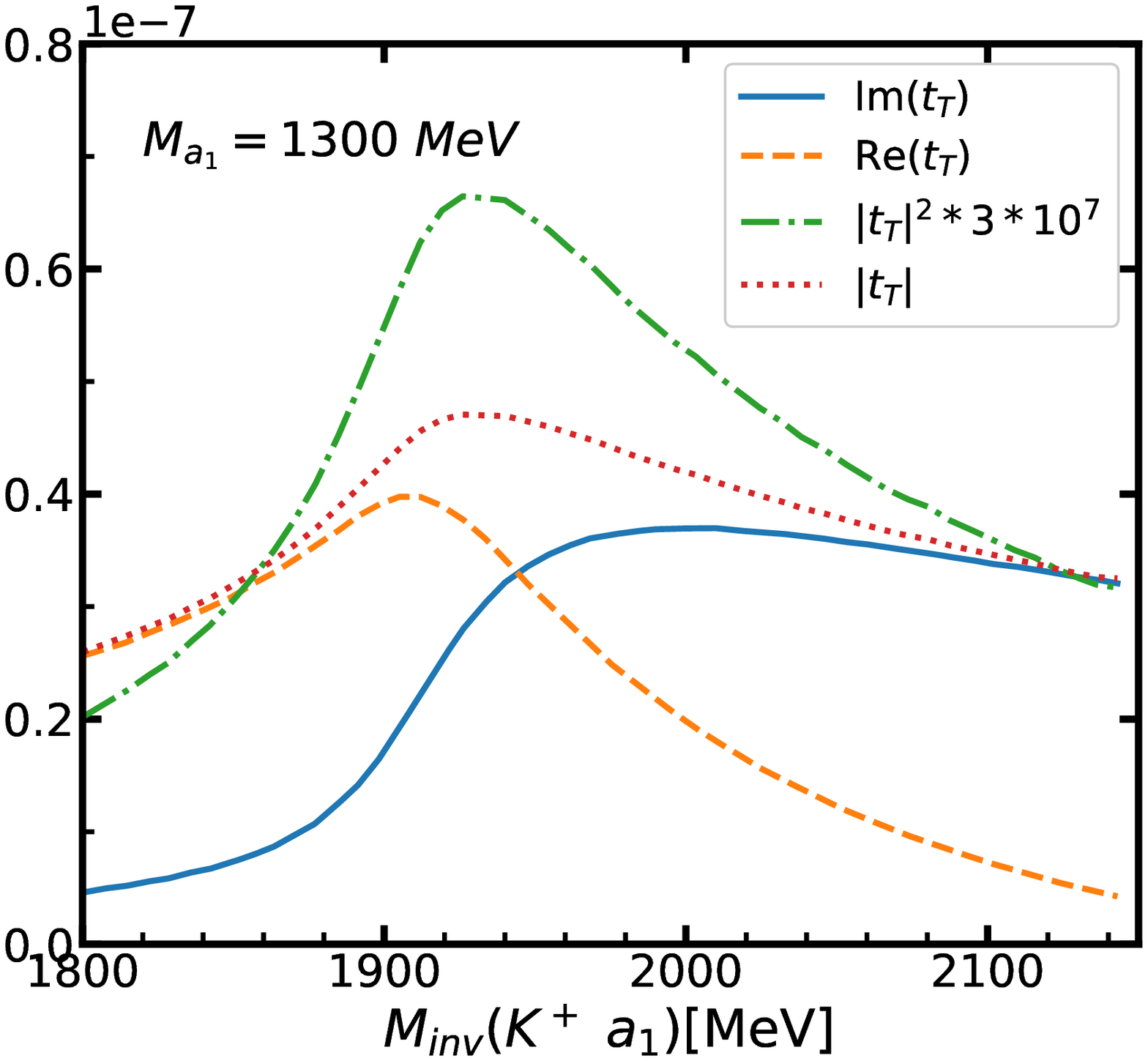}
\label{a}
}
\subfigure[]
{
\includegraphics[scale=0.35]{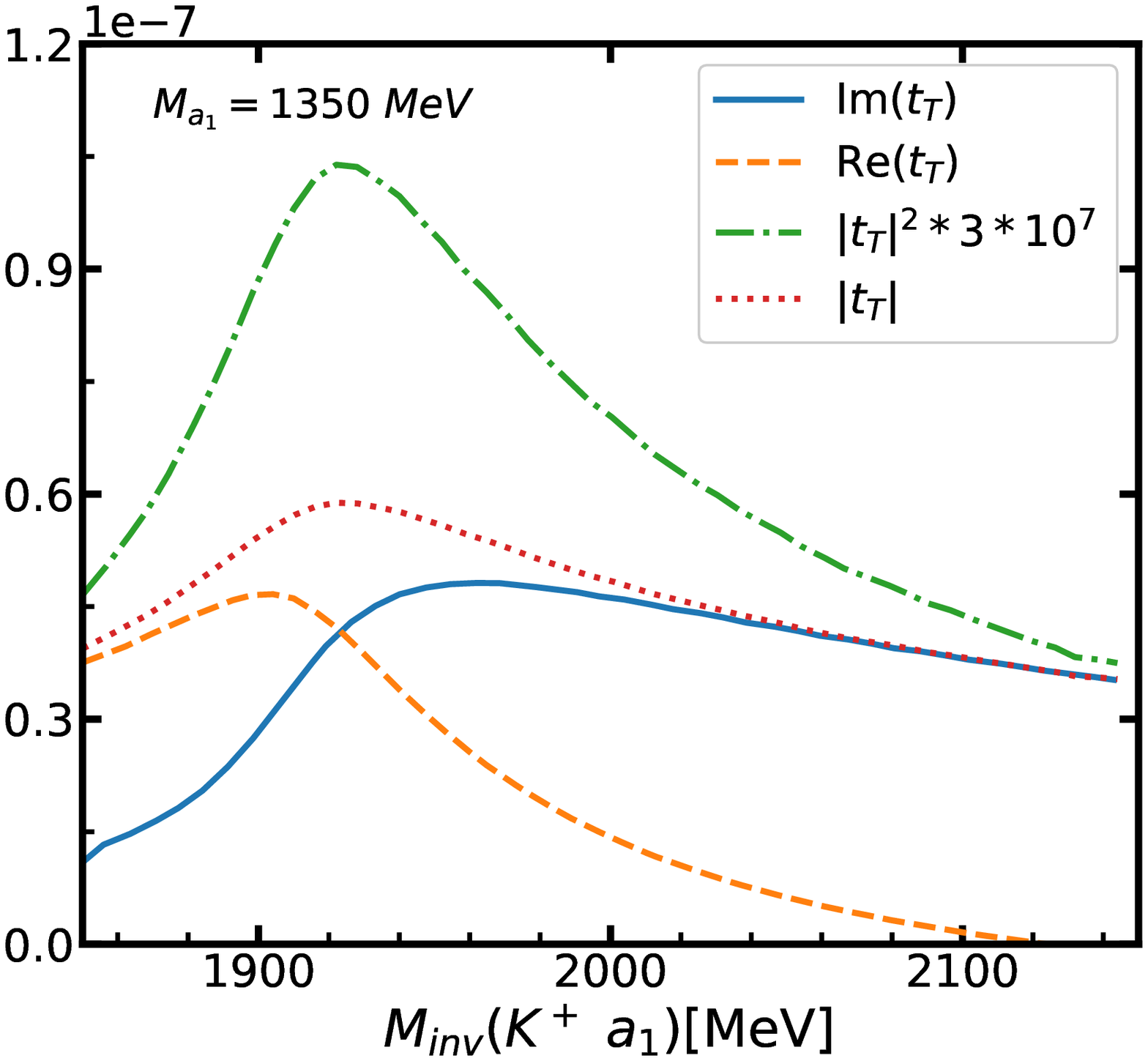}
\label{b}
}
\subfigure[]
{
\includegraphics[scale=0.35]{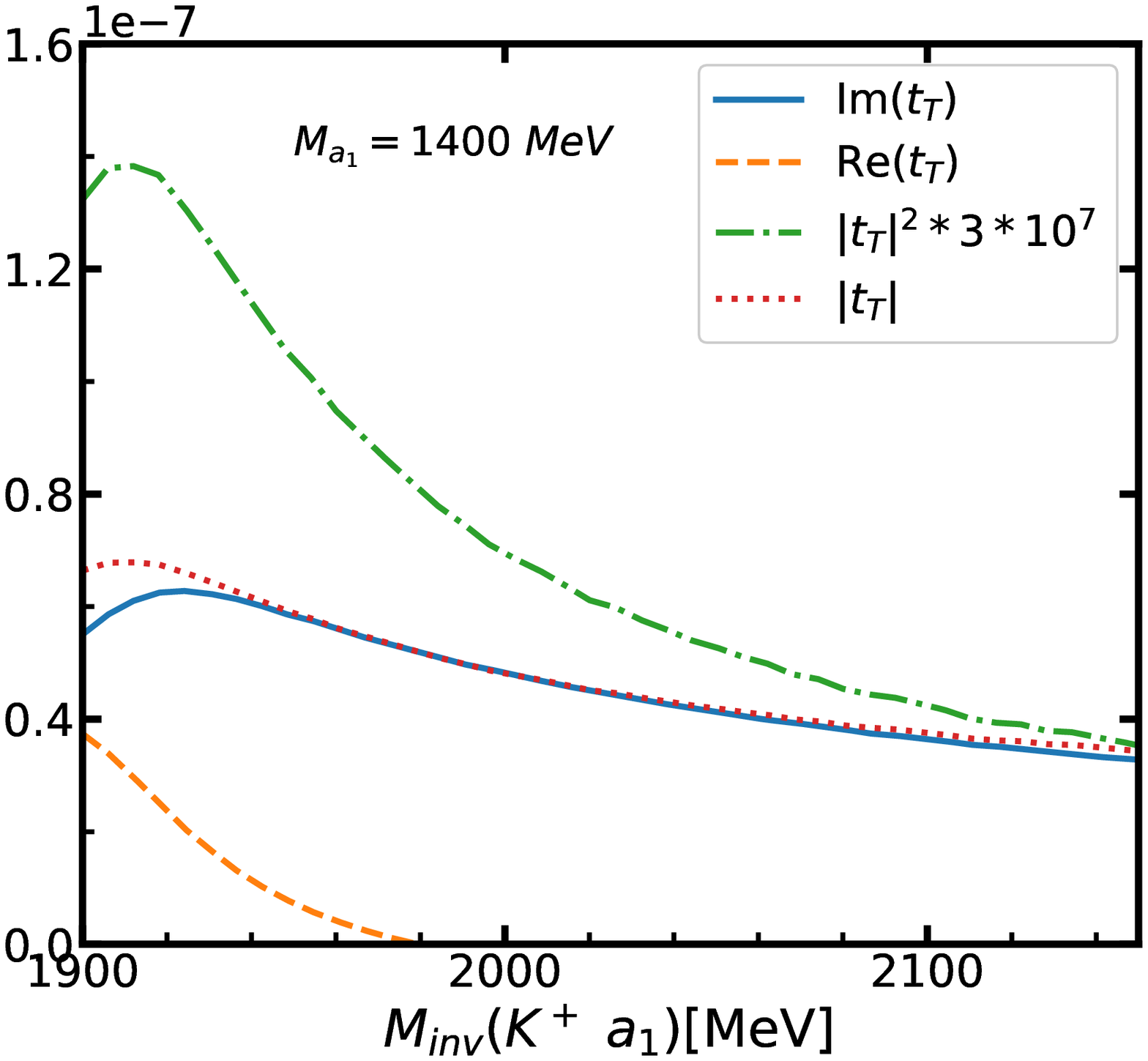}
\label{c}
}
\end{center}
\vspace{-0.7cm}
\caption{Triangle amplitude $t_T$ as a function of $M_{\rm inv}(K^+ a_1)$ for (a) $M_{\rm inv}(\pi^+ \rho^-)=1300 MeV$, (b) $M_{\rm inv}(\pi^+ \rho^-)=1350 MeV$ and (c) $M_{\rm inv}(\pi^+ \rho^-)=1400 MeV$. $|t_T|^2,|t_T|,\text{Re}(t_T)$ and $\text{Im}(t_T)$ are plotted using the green, red, yellow and blue curves, respectively.}
\label{Fig2}
\end{figure*}
\begin{figure*}[htp!]
\begin{center}
\includegraphics[scale=0.35]{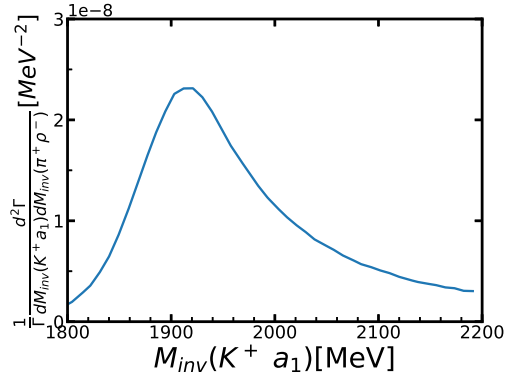}
\end{center}
\vspace{-0.7cm}
\caption{The differential branching	ratio $\displaystyle \frac{1}{\Gamma}\frac{d^2 \Gamma}{dM_{\rm inv}(K^+ a_1)dM_{\rm inv}(\rho \pi)}$ described as in Eq.~\eqref{Gamma} as a function of $M_{\rm inv}(K^+ a_1)$. The integration range of $M_{\rm inv}(\pi^+ \rho^-)$ is given by the text.}
\label{Fig3}
\end{figure*}
\begin{figure*}[htp!]
\begin{center}
\includegraphics[scale=0.35]{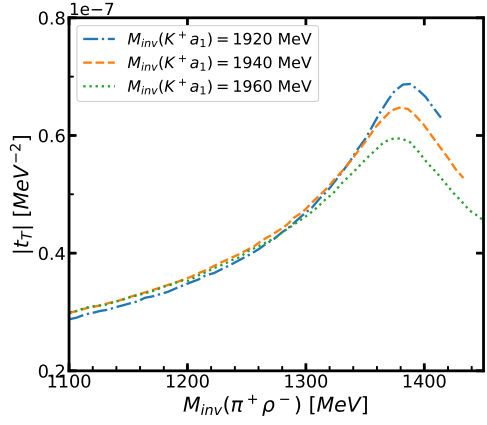}
\end{center}
\vspace{-0.7cm}
\caption{Triangle amplitude $|t_T|$ as a function of $M_{\rm inv}(\pi^+ \rho^-)$ for $M_{\rm inv}(K^+ a_1)$=1920, 1940 and 1960 MeV.}
\label{Fig4}
\end{figure*}
\begin{figure*}[htp!]
\begin{center}
\includegraphics[scale=0.35]{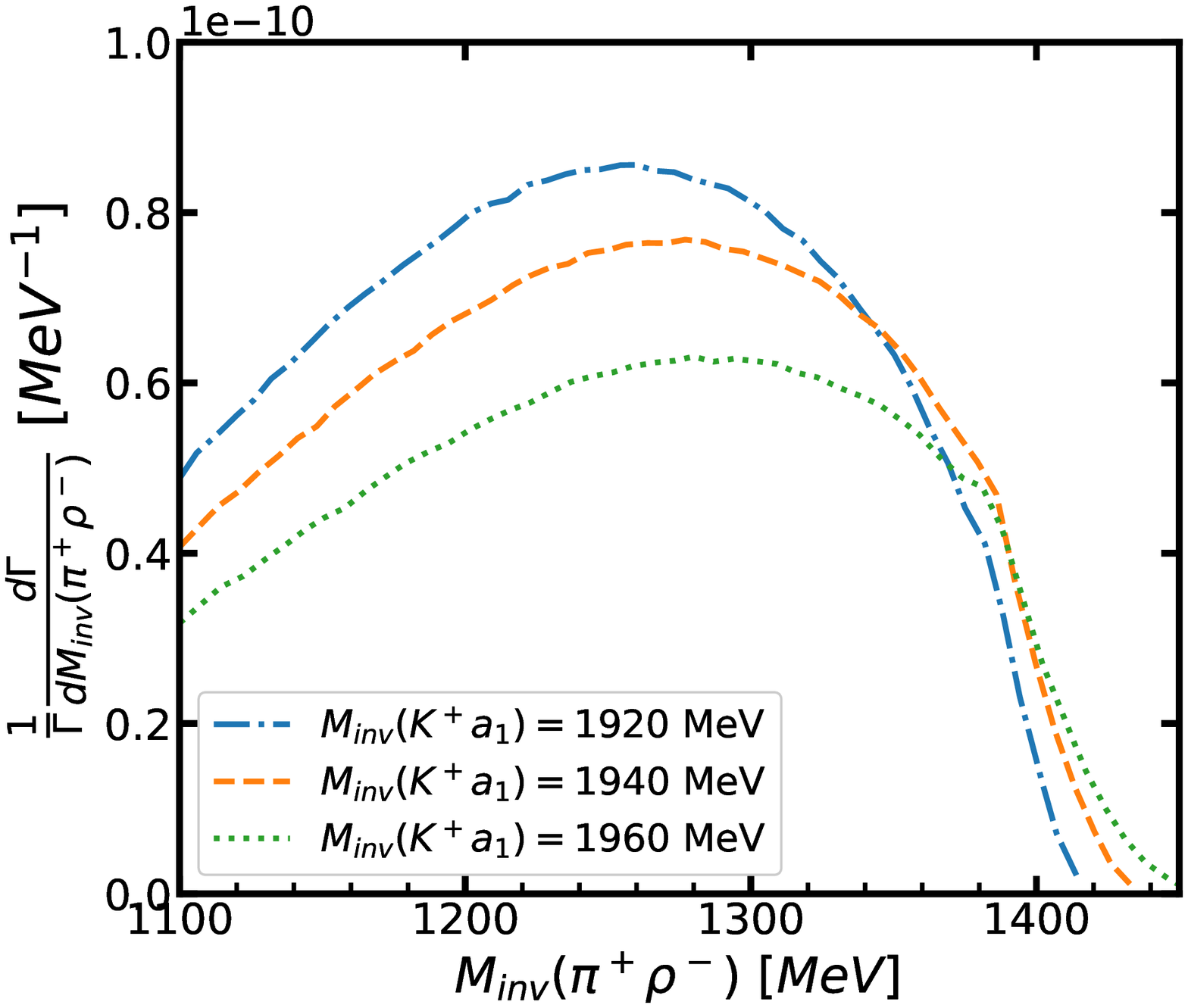}
\end{center}
\vspace{-0.7cm}
\caption{The differential branching	ratio $\displaystyle \frac{1}{\Gamma}\frac{d^2 \Gamma}{dM_{\rm inv}(K^+ a_1)dM_{\rm inv}(\pi^+ \rho^-)}$ described as in Eq.~\eqref{Gamma} as a function of $M_{\rm inv}(\pi^+ \rho^-)$ for $M_{\rm inv}(K^+ a_1)$=1920, 1940 and 1960 MeV.}
\label{Fig5}
\end{figure*}

At the beginning, we present in Figs.\ref{a}, \ref{b} and \ref{c} the absolute value, square of the absolute value, imaginary part, and real part of the triangle loop amplitude $t_T$ in Eq.\eqref{tT} as functions of the $K^+ a_1$ invariant mass, where the invariant masses of final $\pi^+ \rho^-$ states are taken as 1300, 1350 and 1400 MeV, respectively. 
We focus on Fig.\ref{c} obtained by taking the invariant masses of $\pi^+ \rho^-$ a little larger than the $K^- K^{*+}$ threshold 1385 MeV first. Note that in this case the smallest value of $M_{\rm inv}(K^- a_1)$ is about 1900 MeV (nearby the $K^+ a_1$ threshold). From this diagram we can see that there is a peak in $|t_T|^2$ located at around 1920 MeV with a width of 100 MeV. This width mainly originate from the width of the vector propagators $K^*$ and $\phi$, which is basically consistent with the prediction 96 MeV. It is also found from Figs.\ref{a}, \ref{b} and \ref{c} that as the chosen mass of $a_1(1260)$ becomes larger, the width of the peak in $|t_T|^2$ becomes larger.
%There is a gap between the previous predicted width of the triangle singularity (62 MeV) and 150 MeV. This is because %the latter is obtained with $a_1(1260)$ the mass $1230$MeV which is a bit lower than the $K^* K$ threshold. One can %easily find by solving Eq.\eqref{eq1} that as the mass of $a_1(1260)$ slightly smaller than the $K^* K$ threshold, the %"width" of the corresponding triangle singularity becomes larger. 
A bump in ${\rm Im}(t_T)$ can be found nearby 1920 MeV related with the triangle singularity, which has been addressed at Refs.~\cite{Sakai:2017hpg,Dai:2018hqb}. After comparing with Figs.\ref{a} and \ref{b}, we find that the strengths of the absolute value, square of the absolute value, imaginary part, and real part of the triangle loop amplitude $t_T$ become smaller as the $M_{a_1}$ get smaller. This suggests all these quantities can obtain sizeable enhancement when $M_{a_1}$ is close to the $K^- K^{*+}$ threshold. Also, the position of the peak in $|t_T|^2$ leaves basically unchanged due to the triangle mechanism. We observe a broad peak in ${\rm Re}(t_T)$ nearby the $K^* \phi$ threshold which has been suggested and discussed in Refs.~\cite{Sakai:2017hpg,Dai:2018hqb}. Moreover, the primary bump in Fig.\ref{c} has converted into broad bumps in Figs.\ref{a} and \ref{b} on account of the potential deviation from the $K^- K^{*+}$ threshold required by the triangle mechanism.

As shown in Fig.\ref{Fig3}, we plot the differential branching ratio of the underlying decay process $\frac{1}{\Gamma}\frac{d^2 \Gamma}{dM_{\rm inv}(K^+ a_1) dM_{\rm inv}(\pi^+\rho^-)}$ defined in Eq.~\eqref{Gamma}, where the $M_{\rm inv}(\pi^+ \rho^-)$ has been integrated from $(m_{\pi^+}+m_{\rho^-})$ to $(M_{\rm inv}(K^+ a_1)-m_{K^-})$. There is a clear peak around 1920 MeV as predicted by the triangle mechanism. The strength of the differential branching ratio can reach $2.4\times 10^{-8}$MeV$^{-2}$. What's more, the upper part of the invariant mass distribution drops more slowly than the lower part because of the polarized factor in Eq.~\eqref{pol}, which produces large contribution when $M_{\rm inv}(\pi^+ \rho^-)$ is large.
We also plot the triangle amplitude $|t_T|$ in Eq.~\eqref{tT} as a function of $M_{\rm inv}(\pi^+ \rho^-)$ with $M_{\rm inv}(K^+ a_1)$ taken by 1920, 1940 and 1960 MeV in Fig.\ref{Fig4}, respectively. There is a peak near 1390 MeV in all these three case which is the direct reflection of the triangle mechanism: to obtain the triangle singularities, one should let the $M_{\rm inv}(\pi^+ \rho^-)$ slightly larger than the $K^- K^{*+}$ threshold. It is desirable to mention that there is a very small reduction on the position of the peak as the $M_{\rm inv}(K^+ a_1)$ increases from 1920 to 1960 MeV. On the other hand, the distribution with the $M_{\rm inv}(K^+ a_1)$ taken as 1920 MeV has the largest strength which is enhanced by the triangle mechanism. As the $M_{\rm inv}(K^+ a_1)$ increases from the position of triangle singularity, the strength becomes lower and the width of the peak gets larger.

Next, we show the differential branching ratio $\displaystyle \frac{1}{\Gamma}\frac{d \Gamma}{dM_{\rm inv}( \pi^+ \rho^-)}$ described as in Eq.~\eqref{Gamma} as a function of $M_{\rm inv}(\pi^+ \rho^-)$ for $M_{\rm inv}(K^+ a_1)$=1920, 1940 and 1960 MeV. We find that the $M_{\rm inv}(\pi^+ \rho^-)$ distribution around 1920 MeV has the largest strength over the three cases, which is a natural result of the triangle mechanism. Note that the position of the peak in this three cases increases as the $M_{\rm inv}(\pi^+ \rho^-)$ increases. The peaks of all three distributions deviate from the $K^- K^{*+}$ threshold due to the contribution coming from the polarized factor involving $k^2$ in Eq.~\eqref{pol}.

Finally, we integrate out two invariant masses in Eq.~\eqref{Gamma} in order to obtain the branching ratio of the total decay process $J/\psi \to K^- K^+ a_1(1260)$. The integration range of $M_{\rm inv}(K^+ a_1)$ is $(m_{K^+}+m_{a_1},m_{J/\psi}-m_{K^+}-m_{a_1})$, while those for $M_{\rm inv}(\pi^+ \rho^{-})$ is $(m_{\rho^-}+m_{\pi^+}, M_{\rm inv}(K^+ a_1)-m_{K^-})$. We find
\begin{equation}
\begin{aligned}
\text{Br}(J/\psi \to K^- K^+ a_1(1260),a_1\to \pi^+ \rho^-)=4.033 \times 10^{-6},
\end{aligned}
\end{equation}
and then one can easily obtain
\begin{equation}
\begin{aligned}
\text{Br}(J/\psi \to K^- K^+ a_1(1260),a_1\to \pi \rho)=1.210 \times 10^{-5}.
\end{aligned}
\end{equation}
In addition, we obtain the decay branching ratio of $J/\psi \to K^- K^+ a_1(1260)$ by using Eq.~\eqref{gamma}
\begin{equation}
\begin{aligned}
\text{Br}(J/\psi \to K^- K^+ a_1(1260))=3.501 \times 10^{-5}.
\end{aligned}
\end{equation}
These rates are accessible at BESIII within the observation capability.

\section{Conclusion}
\label{IV}

The present study was designed to determine the effect of triangle mechanism of the dacay process of $J/\psi \to K^- K^+ a_1(1260)$. The results of this investigation show that there is a triangle singularity around 1920 MeV for the invariant mass $M_{\rm inv}(K^+ a_1)$.
The strength of the differential branching ratio $\displaystyle \frac{1}{\Gamma}\frac{d^2 \Gamma}{dM_{\rm inv}(K^+ a_1)dM_{\rm inv}(\rho^- \pi^+)}$ reaches 2.4$\times 10^{-8}$ MeV$^{-1}$. We have applied the experimental data of the branching ratio of the decay $J/\psi \to \bar{K} K^{*} \phi$ to determine the coupling strength of the $J/\psi \bar{K} K^* \phi$ vertex. 
We also evaluate the triangle amplitude $|t_T|$ and the differential branching ratio $\displaystyle \frac{1}{\Gamma}\frac{d^2 \Gamma}{dM_{\rm inv}(K^+ a_1)dM_{\rm inv}(\rho^- \pi^+)}$ as functions of $M_{\rm inv}(\pi^+ \rho^-)$. There are deviations of all three distributions in the latter from the $K^- K^{*+}$ threshold on account of the contribution coming from the polarized factor involving $k^2$.
We hope that the future the LHCb, Belle-II and BESIII experimental data will focus on the $J/\psi \to K^- K^+ a_1(1260)$ process and clarify the role played by triangle singularities in this decay process, which can provide valuable information for the low lying axial-vector mesons $a_1(1260)$.

\begin{acknowledgments}
The authors thank professor Jujun Xie for his patient guidance. Hao Sun is supported by the National Natural Science Foundation of China (Grant No.12075043).
\end{acknowledgments}

\bibliography{ref}

\end{document}